# Density Fluctuations in Uniform Quantum Gases*

J. Bosse[1], K.N. Pathak[2] and G.S. Singh[3]

[1]*Institute for Theoretical Physics, Freie Universität Berlin, Germany*
[2]*Department of Physics, Panjab University, Chandigarh, India*
[3]*Department of Physics, Indian Institute of Technology, Roorkee, India*

**Abstract.** Analytical expressions are given for the static structure factor S(k) and the pair correlation function g(r) for uniform ideal Bose-Einstein and Fermi-Dirac gases for all temperatures. In the vicinity of Bose Einstein condensation (BEC) temperature, g(r) becomes long ranged and remains so in the condensed phase. In the dilute gas limit, g(r) of bosons & fermions do not coincide with Maxwell-Boltzmann gas but exhibit bunching & anti-bunching effect respectively. The width of these functions depends on the temperature and is scaled as √ (inverse atomic mass). Our numerical results provide the precise quantitative values of suppression/increase (antibunching and bunching) of the density fluctuations at small distances in ideal quantum gases in qualitative agreement with the experimental observation for almost non-trapped dilute gases.

## INTRODUCTION

The study of the pair correlation function and the structure factor of fluid matter have been of interest both experimentally and theoretically since long and have yielded important information on the structure of the systems. However, the study of these correlations in cold atomic gases is getting due attention only recently [1-5].

## THEORETICAL PROCEDURE

The (dimensionless) static structure factor $S(\vec{q})$ for a system containing N particles in volume V is defined as

$$S(\vec{q}) = \frac{1}{N} <\delta\rho_{\vec{q}} \delta\rho_{\vec{q}}^\dagger> \quad (1a)$$

$$= \frac{1}{N} \sum_{\vec{k},\vec{k}',\sigma,\sigma'} <\delta(a_{\vec{k},\sigma}^\dagger a_{\vec{k}+\vec{q},\sigma})\delta(a_{\vec{k}'+\vec{q},\sigma'}^\dagger a_{\vec{k}',\sigma'})>, \quad (1b)$$

where $\delta\rho_{\vec{q}} = \rho_{\vec{q}} - <\rho_{\vec{q}}>$ is the operator of density fluctuation of wave vector $\vec{q}$, angular brackets denote the average over a grand canonical ensemble, and the particle-number density $\rho_{\vec{q}} = \sum_{j=1}^{N} \exp(-i\vec{q}\cdot\vec{r}_j)$ has been expressed in second quantized form as $\rho_{\vec{q}} = \sum_{\vec{k}\sigma} a_{\vec{k}\sigma}^+ a_{(\vec{k}+\vec{q})\sigma}$. In Eq.(1), $\vec{q}=0$ corresponds to uniform constant density n=N/V (no fluctuations).

For a non-interacting system described by sum of single particle Hamiltonian, Eq. (1) will reduce to [6],

$$S(\vec{q}) = 1 + \frac{\eta N}{g_s} \sum_{\vec{k}} C_{\vec{k}} C_{\vec{k}+\vec{q}} \quad , \quad q \neq 0 \quad (2)$$

with,

$C_{\vec{k}} = \dfrac{g_s}{(e^{\beta(\varepsilon_k-\mu)} - \eta)N}$ is the average number of particles occupying the quantum state $|\vec{k}\rangle$, $g_s = (2s+1)$ denotes the spin multiplicity & $\sum_{\vec{k}} C_{\vec{k}} = 1$.

In Eq. (2) $\eta = 1(-1)$ for bosons (fermions), $\beta = 1/(k_B T)$, $\varepsilon_k = \hbar^2 k^2/(2m)$ & $\mu$ is chemical potential. For the homogenous and isotropic system of free particles considered here, taking into account explicitly the condensed state of bosons, Eq. (2) further simplifies,

$$S(\vec{q}) = 1 + \frac{\eta N}{g_s}[2C_0 C_q + \sum_{\substack{\vec{k}\neq 0 \\ \vec{q}\neq 0}} C_{\vec{k}} C_{|\vec{k}+\vec{q}|}], \quad (3)$$

where, $C_0 = \delta_{\eta,1}\theta(1-t)(1-t^{3/2})$ is the fraction of particles condensed in the zero momentum state & $t = T/T_c$, $T_c$ being BEC critical temperature. Here S(q) depends only on $q=|\vec{q}|$. Eq. (3) can be simplified to give,

$$S(q) = 1 + \frac{2C_0}{(e^{\beta\varepsilon_q}-1)} + \frac{g_s}{n\lambda^3}\sum_{i,p=1}^{\infty}\frac{e^{\left[\beta\mu(i+p)-\frac{\beta\varepsilon_q ip}{i+p}\right]}}{(i+p)^{3/2}}. \quad (4a)$$

Using Eq. (4a) we obtain for $q>0$ (but can tend to zero in the limiting sense)

$$S(q) = 1 + \frac{2C_0}{(e^{\beta\varepsilon_q}-1)} + \frac{t^{3/2}g_s}{\zeta(3/2)}\sum_{i,p=1}^{\infty}\frac{e^{\frac{-ip\beta\varepsilon_q}{(i+p)}}}{(i+p)^{3/2}}, \quad (4b)$$

$$= 1 + \frac{t^{3/2}g_s}{\zeta(3/2)}\sum_{i,p=1}^{\infty}\frac{e^{(i+p)\beta\mu - \frac{ip\beta\varepsilon_q}{(i+p)}}}{(i+p)^{3/2}}, \quad (4c)$$

for $t<1$ and $t\geq 1$ respectively. Further, for $t\gg 1$ and $\beta\mu\to -\infty$, we get

$$S(q) = 1 + \frac{\eta}{g_s}\frac{n\lambda^3}{2\sqrt{2}}e^{-q^2\lambda^2/(8\pi)}$$

$$= 1 + \frac{\eta}{g_s}\frac{\zeta(3/2)}{(2t)^{3/2}}e^{-\varepsilon_q/(2k_BT_c t)} \quad (4d)$$

Using Eq.(4a) the pair-correlation function g(r) is obtained through the Fourier transform(FT) as,

$$g(r) = 1 + \frac{1}{2\pi^2 nr}\int_0^{\infty}q\sin(qr)[S(q)-1]dq \quad (5)$$

On further simplification we obtain the result [7],

$$g(r) = 1 + \frac{\eta}{g_s}[F(r)^2 - C_0^2], \quad (6)$$

where

$$F(r) = C_0 + \frac{g_s}{n\lambda^3}\sum_{p=1}^{\infty}\frac{e^{[\beta\mu p-\pi r^2/(p\lambda^2)]}}{p^{3/2}}. \quad (7)$$

It can be seen that the last term in Eq.(7) can also be represented by an integral given below.

$$F(r) = C_0 + \frac{\eta g_s}{2\pi^2 nr}\int_0^{\infty}dk\frac{k\sin(kr)}{e^{\beta(\varepsilon_k-\mu)}-\eta}. \quad (8)$$

It is noted that the present derivation provides the analytical expression for boson S(q) for all T, the FT of which gives g(r). It seems to us that the result given in Eq.(4a) is new. It has also been checked that one can also obtain the same result for S(q) by first calculating g(r) then taking its FT. Further Eq.(7) simplifies,

$$F(r) = 1 - t^{3/2} + \frac{t^{3/2}}{\zeta(3/2)}\sum_{p=1}^{\infty}\frac{e^{\frac{-\pi r^2}{p\lambda^2}}}{p^{3/2}}, \quad t\leq 1$$

$$= \frac{t^{3/2}}{\zeta(3/2)}\sum_{p=1}^{\infty}\frac{e^{p\beta\mu - \frac{\pi r^2}{p\lambda^2}}}{p^{3/2}} \quad t\geq 1 \quad (9)$$

For high $T, \beta\mu\to -\infty$, we find $F(r) = e^{-\pi r^2/\lambda^2}$ & $g(r) = 1 + \frac{\eta}{g_s}e^{-\frac{2\pi r^2}{\lambda^2}}$. The difference in the approach of S(q) & g(r) to unity for bosons & fermions at high T is to be noted. The chemical potential $\mu$ is determined from the normalization condition,

$$1 = \frac{g_s}{n\lambda^3}\frac{Li_{3/2}(\eta e^{\beta\mu})}{\eta}, \quad (10)$$

$Li_j(x) = \sum_{k=1}^{\infty}\frac{x^k}{k^j}$ is the polylog function and $\lambda = \sqrt{2\pi\hbar^2\beta/m}$ the de-Broglie wavelength.

For a fermi gas, calculations proceed differently due to chemical potential being positive up to some temperature $T_0$ & negative thereafter. Results are given elsewhere [7]. The function g(r) of a bose gas is calculated either from Eqs. (6) & (7) or (8), using $\mu(T)$ from Eq. (10). It is displayed in Fig. 1(Color online) for the temperatures t=0.3(blue), 0.95(green), 1.05(pink), 5.00(red) respectively, corresponding to $k_BT/\varepsilon_u \approx 0.0327, 0.1036, 0.1145, 0.5451$, where units used are

$$k_u = 2(\frac{6\pi^2 n}{2s+1})^{\frac{1}{3}} = 2k_F \quad ; \varepsilon_u = \frac{\hbar^2 k_u^2}{2m} = 4\varepsilon_F \quad (11)$$

The BEC critical temperature $T_c$ is

$$T_c \approx \frac{0.109\,\varepsilon_u}{k_B} \quad or \quad T_c = 0.436 T_F,$$

where $T_F, \varepsilon_F$ & $k_F$ denote Fermi temperature, Fermi energy and Fermi wave number, respectively.

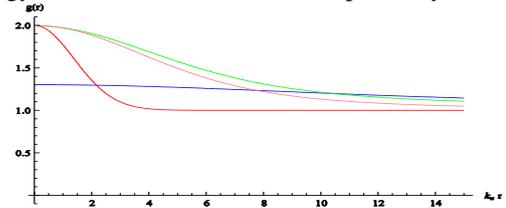

**Figure 1.** g(r) for s=0 bosons.

The increase in g(r)-1, density fluctuation (bose pile) for small distances from a value 0 to 1 is in qualitative agreement with experiments for a degenerate dilute bose gas. It is noted that g(0) =1 at T=0 and rises with t & attains a value 2 at t=1& remains the same for t>1. This has also been found to be true for trapped gas. Details of this & other results for g(r) for trapped gases will be reported elsewhere[8]. The "bose pile" observed in g(r) for small distances r is associated with a temperature–dependent *attractive* effective pair–potential,

$$\varphi(r) = -k_BT \ln[g(r)] \Leftrightarrow g(r) = e^{\frac{-\varphi(r)}{k_BT}} \quad (12)$$

which is displayed in Fig.2 (Color online) for the above set of temperatures.

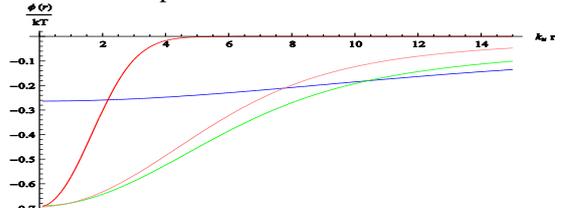

**Figure 2.** Effective Potential for s=0 bosons.

Note that the range of the attractive boson–boson potential *increases* as T decreases towards $T_c$. Below $T_c$, it will continue to increase, which is an artifact of the non–interacting system.

The function g(r) of a Fermi gas calculated from Eq.(6) & (7) or (8), using $\mu(T)$ from Eq.(10) is displayed in Fig. 3 (color online) for the above set of temperatures, which may be expressed in terms of the Fermi temperature as $t \approx$ 0.1308(blue), 0.4143(green), 0.4579(pink), 2.180(red).

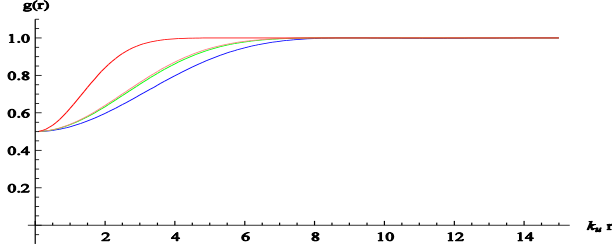

**Figure 3:** g(r) for s=1/2 Fermions.

Again the decrease in the density fluctuation from a value 1 to 0.5 for small distances is in qualitative agreement with experiments. The "Fermi hole" observed in g(r) for small distances r is associated with a temperature–dependent *repulsive* effective pair–potential,

$$\varphi(r) = -k_B T \ln[g(r)] \Leftrightarrow g(r) = e^{\frac{-\varphi(r)}{k_B T}} \quad (13)$$

and is shown in Fig. 4(Color online) for the above set of temperatures.

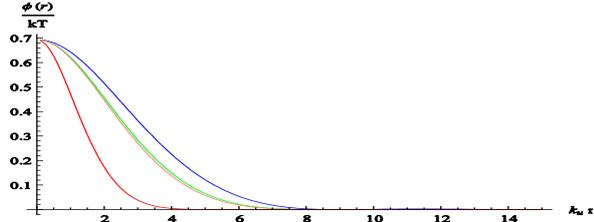

**Figure 4.** Effective Potential for s=1/2 fermions.

It is noted that the range of the repulsive Fermion–Fermion potential also *increases* as T decreases. However, in contrast to the ideal bose gas, the range does not increase infinitely but terminates. It is a short ranged force.

The strength of the effective potential at $r=0$ is

$$\frac{\varphi(r=0)}{k_B T} = -\ln[1 - \frac{1}{2s+1}],$$
$$= -\ln[1 + \frac{1}{(2s+1)}], \quad (14)$$

for fermi and bose gases, respectively.

S(q) can be calculated either from Eq. (4a) involving double sum or using the dynamical calculation of response function & fluctuation dissipation theorem according to the expression,

$$S(q) = \int_{-\infty}^{\infty} d\omega \; S(q,\omega) = \frac{\hbar}{n\pi} \int_0^{\infty} d\omega \coth(\frac{\beta\hbar\omega}{2}) \chi''(q,\omega), \quad (15)$$

The imaginary part of density response function $\chi''(q,\omega)$, analytical expression for which already given in reference [9]. The latter way of calculation is much faster than evaluation of the infinite double sum. The numerical results from both the procedure for the static structure factor for spin-zero bose gas at temperatures $t$=0.3(blue), 0.95 (green), 1.05 (pink), 5.00 (red) are shown in Fig 5. Full lines & dots are results obtained from Eq. (15) & Eq. (4a) respectively. Blue and green curves approach ∞ for $q \to 0$ reflecting the diverging compressibility of the ideal bose gas. Both the procedure gives identical results except near $T_c$ where results from the Eq. (15) are more reliable.

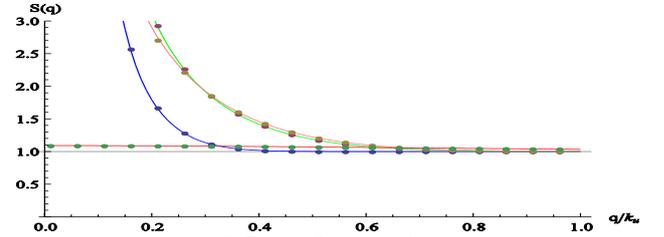

**Figure 5.** Boson Structure factor.

Now, the static structure factor for quantum gases at temperatures $t$ =0.3 (blue), 0.95 (green), 1.05 (pink), 5.00 (red) for bosons ($S(q) \geq 1$) and fermions ($0 \leq S(q) \leq 1$) are presented in Fig. 6 (Color online) to have the quantitative estimates. It is known and can also be seen from Fig 6 (as well from Eq. (4a)) that boson structure factor diverges for q→0. This can be repaired by including the interaction [9]

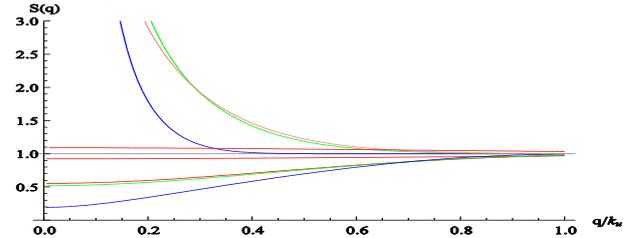

**Figure 6.** Structure factor for bose & fermi gases.

We also compare the Bogoliubov structure factor with that obtained for the ideal gas to see the difference in their quantitative behavior. For this we need to use the result [10]

$$S(q) = \frac{\hbar^2 q^2}{2m\omega_q} \coth(\frac{\beta\hbar\omega_q}{2}), \quad (16)$$

where

$$\omega_q = \sqrt{(\frac{\hbar q^2}{2m})^2 + (cq)^2}, \quad (17)$$

is the Bogoliubov excitation frequency and it tends to $c\,q$ for q→0, c being the isothermal sound velocity.

We also note the exact result $\lim_{q\to 0} S(q) = n\, k_B\, T\, \kappa_T$, implying that $S(0) \to 0$ for $T=0$. In Fig 7, $S(q)$ of BEC for various reduced temperatures $T_r = k_B T/m\, c^2 = 0$ (dashed), 0.2 (blue), 1 (green), 2 (pink), 3 (red) is shown. Although wave number & temperature scale are different here from that used in Figs. 5 & 6, one notices the qualitative differences in the behavior of the structure factor due to interactions. Especially, the boson structure factor will be smaller than 1 for large $q$, similar to the fermion structure factor. For very low temperatures (blue curve in Fig.7), the boson structure factor of the interacting bose gas will be very different from that of an ideal bose gas (blue boson curve in Fig.6), while it is very similar to that of an ideal Fermi gas (blue Fermi curve in Fig.6).

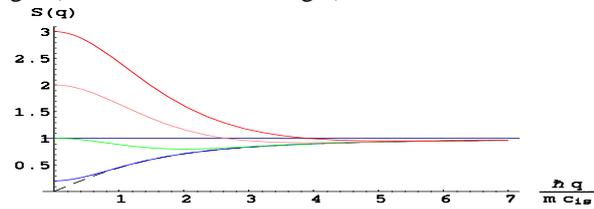

**Figure 7**. Static Structure factor of B E Condensate.

In Fig. (8), we compare $S(q)$ of bose gas obtained from Eq.(16) and Eqs.(15) or (4a) at temperatures $t=0.1$ (blue), 0.2 (violet), 0.3 (brown), 0.4 (green). Dots represent results Eq. (4a). Dashed lines result from Eq. (16) for $2mc/(\hbar k_u) = 0.025$. It is noted from the Fig. that this value of c reproduces the ideal bose gas structure factors for $q \geq 0.11\, k_u$ at low temperatures.

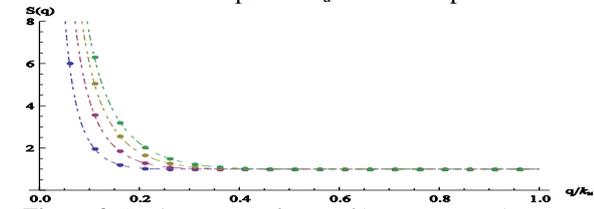

**Figure 8.** Static structure factor of homogeneous bose gas.

## DISCUSSION

In this paper we have obtained S(q) & g(r) of homogeneous ideal bose gas at all temperature. The first one seems at least to us a new result while later has also been discussed by Landau and Lifshitz [11] but have not made any numerical estimates. Particle-counting experiments [12,13] on cold atomic fermi and bose gases released from a trap give the first observation of quantum statistical interference effects which has been observed originally by Hanbury-Brown and Twiss on photons about 50 years ago [14]. Jeltes et al. [12] experimentally measured the normalized pair-correlation functions for the two isotopes of $^3$He and $^4$He under same conditions and they found the correlation lengths in $^3$He, $^4$He scale as square root of inverse of their atomic mass respectively. This is in agreement with our theoretical results for non-interacting gases. It is not possible to make point to point comparison of their result with our calculation at this stage, however, we find that the ratio,

$$r(T) = \frac{g_B(r=0)}{g_F(r=0)}\; at\;\; T < T_c$$

The ratio varies between

$$r(T=0) = 1 + \frac{1}{(2 S_F)}$$
$$= 2 \text{ if } S_F=1/2$$
$$= 1.33 \text{ if } S_F=3/2$$

and

$$r(T \geq T_c) = \frac{(S_B+1)(2S_F+1)}{S_F(2S_B+1)}$$
$$= 4, \text{ if } S_B=0, S_F=1/2$$
$$= 1.78, \text{ if } S_B=1, S_F=3/2.$$

This ratio in the experiments of Jeltes et al. is 1.1 suggesting that our theoretical predictions are within the ballpark of experimental values.

## ACKNOWLEDGMENT

J Bosse and K N Pathak gratefully acknowledge financial support from the Alexander von Humboldt Foundation. We also acknowledge help of Renu Bala in the final preparation of this manuscript.